%% file: YUYUQI.tex
\documentclass[sigconf]{acmart}

\usepackage{booktabs} 
\usepackage{multirow}
\usepackage{graphicx}
\usepackage{subcaption}
\usepackage{float}
\setcopyright{rightsretained}

\acmDOI{}

\acmISBN{}

\acmConference[]{ACSAC 2018}{December 3-7,2018}{San Juan, Puerto Rico, USA}
\copyrightyear{}

\begin{document}
\title{DeepHTTP: Semantics-Structure Model with Attention for Anomalous HTTP Traffic Detection and Pattern Mining}

\author{Yuqi Yu}
\affiliation{%
  \institution{Beihang University}
  \city{Haidian Qu, Beijing Shi}
  \state{China}
}
\email{yuyuqi_bh@buaa.edu.cn}

\author{Hanbin Yan}
\affiliation{%
  \institution{Beihang University, National Computer Network Emergency Response Technical Team/Coordination Center of China}
  \city{Chaoyang Qu, Beijing Shi}
  \state{China}
} 
\email{yhb@cert.org.cn}

\author{Hongchao Guan}
\affiliation{%
  \institution{Beijing University of Posts and Telecommunications}
  \city{Haidian Qu, Beijing Shi}
  \state{China}
}
\email{ghc@bupt.edu.cn}

\author{Hao Zhou}
\affiliation{%
  \institution{National Computer Network Emergency Response Technical Team/Coordination Center of China}
  \city{Chaoyang Qu, Beijing Shi}
  \state{China}
}
\email{zhh@cert.org.cn}






\begin{abstract}
In the Internet age, cyber-attacks occur frequently with complex types. Traffic generated by access activities can record website status and user request information, which brings a great opportunity for network attack detection. Among diverse network protocols, Hypertext Transfer Protocol (HTTP) is widely used in government, organizations and enterprises. In this work, we propose DeepHTTP, a semantics structure integration model utilizing Bidirectional Long Short-Term Memory (Bi-LSTM) with attention mechanism to model HTTP traffic as a natural language sequence. In addition to extracting traffic content information, we integrate structural information to enhance the generalization capabilities of the model. Moreover, the application of attention mechanism can assist in discovering critical parts of anomalous traffic and further mining attack patterns. Additionally, we demonstrate how to incrementally update the data set and retrain model so that it can be adapted to new anomalous traffic. Extensive experimental evaluations over large traffic data have illustrated that DeepHTTP has outstanding performance in traffic detection and pattern discovery.
\end{abstract}

%
%


\maketitle

\input{samplebody-conf}

\bibliographystyle{ACM-Reference-Format}
\bibliography{sample-bibliography}

\end{document}

%% file: samplebody-conf.tex
\section{Introduction}

As the development of computer technology, network systems and applications get increasingly more complex than ever before. More and more bugs and vulnerabilities appear constantly, which poses a great threat to network security. Attacks initiated by exploiting vulnerabilities are getting increasingly more sophisticated. As a result, many traditional anomaly detection methods based on standard mining methodologies are no longer effective. Thus, how to discover various hidden web attacks is still a topic of great concern in the field of network security.

HTTP, a representative of network protocol, occupies a considerable proportion of the application layer traffic of the Internet. Since HTTP traffic can record website access states and request content, it provides an excellent source of information for anomaly detection \cite{Jamdagni:2010:IDU:1815396.1815669, Tombini:2004:SCA:1038254.1038335, Estevez-Tapiador:2004:MNH:1006340.1006347}. According to the characteristics, existing approaches for anomalous HTTP traffic detection can be roughly divided into two categories: feature distribution based methods \cite{Lakhina:2005:MAU:1090191.1080118, Samant2010Feature} and content-based detection technologies \cite{Swarnkar2016OCPAD}. Additionally, since malicious traffic detection is essentially an imbalanced classification problem, many researches propose anomaly-based detection approaches that generate models merely from the benign network data\cite{Bortolameotti:2017:DDA:3134600.3134605}. With the rapid development of artificial intelligence, deep learning has been widely used in various fields and has a remarkable effect in natural language processing. Recently, deep learning has been successfully applied in anomaly detection\cite{Javaid:2016:DLA:2954721.2954780,DBLP:journals/corr/abs-1803-10769}.

Even though these methods are successful in certain scenarios, they are not universal. The main difficulties and challenges of malicious traffic detection are as follows. First of all, it is difficult to automatically detect hidden anomalous traffic from massive network traffic with lots of noise. How to enhance the generalization ability and robustness of the model is still a critical issue. Secondly, in practical applications, anomaly-based detection model usually has a high false positive rate. Since the model is trained based on normal samples, traffic not present in the training set is likely to be labeled as malicious. This problem undoubtedly increases the workload of manual verification. Last but not least, grasping the characteristics and disciplines of traffic helps to improve defense rules, which is crucial in practical application scenarios. However, the study of pattern mining for malicious traffic is not yet mature. It is remains to be challenging task.

To resolve the above challenges, in this work, we propose DeepHTTP, a data-driven method leveraging the large volumes of traffic data. We assume that the new malicious traffic is similar to the known anomalous traffic in terms of structure or content. Indeed, each HTTP request follows strict architectural standards and language logic, which is similar to natural language. Hence, we treat elements in traffic content as vocabulary in natural language processing and establish models to learn the semantic relationships between them. Meanwhile, we proposes a method for extracting traffic structure information. Specifically, we segment the content of the traffic entries and convert content clips to structural features. The integration of semantics and structural features helps to improve the generalization of the model.

The model we proposed is a combined semantics structure model utilizing Bidirectional Long Short-Term Memory (Bi-LSTM) with attention mechanism. Bi-LSTM can obtain information from past and future states, which allows the model to learn the text characteristics better. Moreover, neural network model with attention mechanism can automatically dig out key information to distinguish different traffic. To discovery the patterns, we cluster samples that have been labeled as malicious and perform pattern mining for each cluster. Since it is a learning-driven approach, we set up a process that can verify and update data efficiently. The model is updated periodically so that it can adapt to new malicious traffic that appears over time. DeepHTTP is a complete framework that can automatically distinguish malicious traffic and mine patterns without prior knowledge.

In summary, we make the following contributions.

We have an in-depth analysis of the types and encoding forms of content of HTTP traffic, then propose an effective structure extraction method.

We adopt an integration framework that combines two attention-based Bi-LSTM models training upon semantics and structure features, separately. Vectors that combine semantic and structural information are ultimately used for classification. The proposed model has superior learning ability and generalization ability.

We propose a novel approach for mining traffic patterns. We use the attentional hidden states of traffic learned by the model as input for clustering. For each cluster, we perform pattern mining based on weight vector obtained from attention model. Experiments prove the effectiveness of our approach.

The rest of this paper is organized as follows. Section 2 gives a summary of the relevant research. Section 3 briefly introduces the system framework and data preprocessing methods, especially the extraction method of the structural characteristics of traffic. The proposed model is introduced in detail in section 4, including the malicious traffic detection model and pattern mining method. We launched a comprehensive experiment to demonstrate the effectiveness of the model. The experimental results are shown in section 5. Section 6 gives the conclusions and future works.
\begin{figure*}[htbp]
\includegraphics[height=5cm, width=12cm]{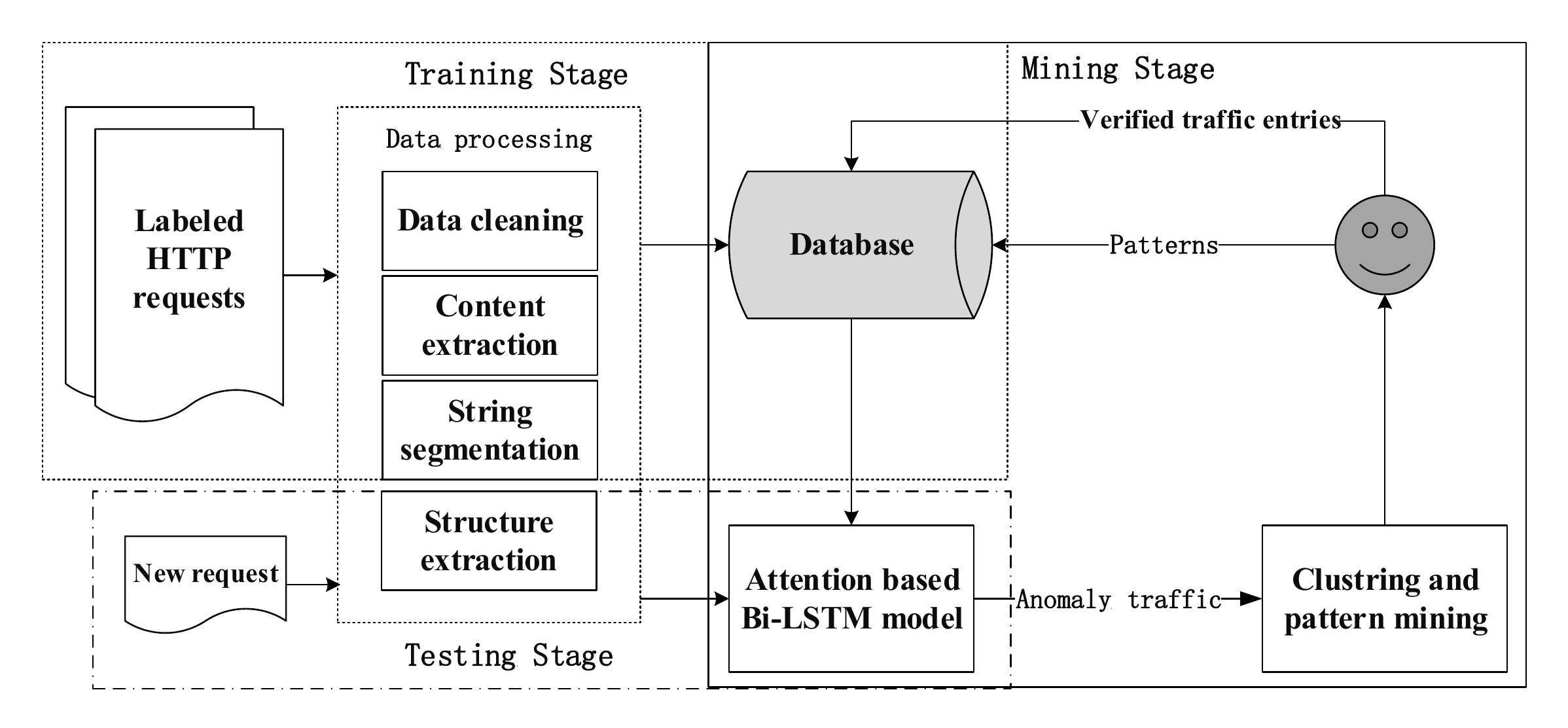}
\caption{DeepHTTP architecture.}
\end{figure*}
\section{Related Work}
In recent years, there are quite a few researches aiming for detecting anomaly traffic and hidden attacks.

Communication traffic contains lots of information which can be used to mine anomaly behaviors. Lakhina et al.\cite{Lakhina:2004:CNA:1028788.1028813} perform a method that fuses information from flow measurements taken throughout a network. Content features are also valuable for detecting anomaly behaviors. Wang et al.\cite{Wang2006Anagram} present Anagram, a content anomaly detector that models a mixture high-order n-grams designed to detect anomalous and "suspicious" network packet payloads. To select the important features from huge feature spaces, Zseby et al.\cite{Iglesias:2015:ANT:2830192.2830260} propose a multi-stage feature selection method using filters and stepwise regression wrappers to deal with feature selection problem for anomaly detection. The methods mentioned above care less about the structure features of communication payloads which are important for distinguishing anomaly attacking behaviors and mining anomaly patterns. In this paper, we put forward structure extraction approach, which can help enhance the ability of detecting anomaly traffic. The structure feature also makes an important role in pattern mining.

There are various methods and models applied in anomaly detection. Some works have been done based on dimensionality reduction \cite{Juvonen:2015:OAD:2839515.2839556, DBLP:journals/corr/BabaieCA14a, Meimei2016PCA, 7224759}. Juvonen and Sipola \cite{Juvonen:2015:OAD:2839515.2839556} propose a framework to find abnormal behaviors from HTTP server logs based on dimensionality reduction. Researchers compare random projection, principal component analysis and diffusion map for anomaly detection. Ringberg et al. \cite{Ringberg:2007:SPT:1269899.1254895} propose a nonparametric hidden Markov model with explicit state duration, which is applied to cluster and scout the HTTP-session processes. This approach analyses the HTTP traffic by session scale, not the specific traffic entries. Additionally, there are also many researches based on traditional methods such as IDS(intrusion detection system and other rule based system) \cite{Xie2010HTTP,Estevez-Tapiador:2004:MNH:1006340.1006347,Jamdagni:2010:IDU:1815396.1815669, ELALFY201455,1250987}. However, these methods rely on feature engineering, which cannot detect unknown types of traffic. In order to solve this problem, researchers began to apply machine learning and deep learning methods to the anomaly detection field\cite{ERFANI2016121, Du:2017:DAD:3133956.3134015, Al2017Hybrid, Andrysiak2014Network}. Among then, solution\cite{Al2017Hybrid} is a new supervised hybrid machine-learning approach for ubiquitous traffic classification based on fuzzy decision trees with attribute selection. There are also quite a few works based on deep learning methods. Erfani et al.\cite{ERFANI2016121} present a hybrid model where an unsupervised DBN is trained to extract generic underlying features, and a one-class SVM is trained from the features learned by the DBN. LSTM model is used for anomaly detection and diagnosis from System Logs\cite{Du:2017:DAD:3133956.3134015}. In this paper, we present Bi-LSTM based on attention mechanism, which has not been applied yet.This model works well in experiments and has good generalization ability in real traffic.

In addition to detecting malicious traffic and attack behaviors, some researches focus on pattern mining of cluster traffic. Most existing methods for traffic pattern recognition and mining are based on clustering algorithms \cite{Chiou:2014:NSM:2661493.2661501, Le2015Novel}. Le et al. \cite{Le2015Novel} propose a framework for collective anomaly detection using a partition clustering technique to detect anomalies based on an empirical analysis of an attack's characteristics. Since information theoretic co-clustering algorithm is advantageous over regular clustering for creating more fine-grained representation of the data, Mohiuddin Ahmed and Abdun Naser Mahmood\cite{10.1007/978-3-319-23802-9_17} extend the co-clustering algorithm by incorporating the ability to handle categorical attributes which augments the detection accuracy of DoS attacks. In addition to clustering algorithm, JT Ren \cite{1041268}conducts research on network-level traffic pattern recognition, and uses PCA and SVM for feature extraction and classification. I. Paredes-Oliva et al. \cite{6193518} build a system based on an elegant combination of frequent item-set mining with decision tree learning to detect anomalies.

In recent years, the attention-based neural network model has become a research hotspot in deep learning, which is widely used in image processing \cite{xu2015show}, speech recognition \cite{DBLP:journals/corr/ChorowskiBSCB15} and healthcare \cite{DBLP:journals/corr/MaCZYSG17}. The application of attention mechanism in the field of natural language processing has also proved to be extremely effective. Luong et al.\cite{DBLP:journals/corr/LuongPM15} first design two novel types of attention-based models for machine translation. Since the attention mechanism can automatically extract important features from raw data, it has been applied to topic classification\cite{10.1007/978-3-319-70096-0_56}, relation Classification\cite{Zhou2016AttentionBasedBL}, and abstract extraction\cite{Ren:2017:LCS:3077136.3080792}. Nevertheless, to the best of our knowledge, there are few works on detecting anomaly HTTP traffic using sequence model and mining patterns based on attention mechanism. Hence, in this paper, we build a model based on attention mechanism, which can get rid of the dependency of artificial extraction features and do well in traffic pattern mining.

\section{Preliminaries}
\subsection{DeepHTTP Architecture and Overview}
The architecture of DeepHTTP shown in Figure 1 consists of three phases and three components. Specifically, it is divided into training stage, test stage, and verification stage. Structure extraction, anomaly traffic detection model and malicious pattern mining module are main parts that run through the entire framework. These parts will be described in detail in later sections.

\subsubsection{Training Stage}
In the training stage, the core task is the construction of the detection model. Specifically, it includes feature engineering and the training of models. We extract valuable information of the traffic after data processing. To characterize traffic, we fuse the content and structural features of traffic as input to the model. Subsequently, the neural network model with attention mechanism will be trained and iterated on a regular basis. To enhance the robustness of the model, we build data sets containing positive and negative samples in different proportions and use cross-validation to train the model. The best model will be saved for actual traffic detection.

\subsubsection{Testing Stage}
In this stage, the trained model is used for anomaly traffic detection. For each new HTTP request, we apply data processing to deal with it. The content and structural features are feed into the trained model for prediction. The intermediate outputs of the model will be reserved and used in the mining phases.

\subsubsection{Mining Stage}
The main works of this phase are to verify the anomalous traffic labeled by the model and to mine malicious patterns. We obtain attentional hidden state from the proposed model as the input of clustering. For each cluster, we extract a small number of samples for manual verification, and then tag all members in the cluster. Simultaneously, we analyze the critical part of each request according to attention weight vector, and then sum up malicious patterns of each cluster. All results are updated to the database.

\subsection{Data Preprocessing}
We focus on HTTP traffic mainly for three reasons: 1) Hypertext Transfer Protocol is an application layer protocol that is widely used for communication between web browsers and web servers. It is used by most web servers. 2) A large majority of web attacks use HTTP, such as Cross-site scripting attack (XSS), SQL injection and so on. These attacks can be well submerged in massive amounts of normal traffic data. 3) Even though payload of traffic can be modified and confused, it is still extremely helpful for relevant research works.

Selecting valuable feature and performing valid data preprocessing are the foundation of data mining. This section introduces the data processing and feature extraction methods in detail.

\subsubsection{Data Cleaning}
To optimize the follow-up detection and analysis, we perform data processing on original HTTP traffic packages captured by monitor software. We parse packages and extract valuable information including request headers and request bodies. The rest of the data processing includes decoding, deleting erroneous and duplicate data, and filling in missing values.

\subsubsection{Content Extraction}
Malicious information of web attacks is usually contained in the parameter value of the request path and the request body (if a POST request is sent). Hence, we extract Uniform Resource Locator (URL) and payload from processed data.

\subsubsection{String Segmentation}
Text vectorization is the key to text mining. Numerous studies use n-grams \cite{Damashek843} to extract the feature of payloads \cite{10.1007/978-3-540-30143-1_11, Kloft:2008:AFS:1456377.1456395, 6945724}. This method can effectively capture the byte frequency distribution and sequence information, but it is easy to cause dimension disaster. To extract crucial parts of a HTTP request, we split the string with special characters as delimiters instead of n-grams. Here is an instance. Suppose the decoded content is: \textit{ "/tienda1/publico/vaciar.jsp <EOS> B2=Vaciar carrito; DROP TABLE usuarios; SELECT * FROM  datos WHERE nombre LIKE"}. The data after string segmentation is denoted as:\textit{"/ tienda1 / publico / vaciar . jsp <EOS>  B2 = Vaciar carrito ; DROP TABLE usuarios ; SELECT * FROM datos WHERE nombre LIKE"}. Words in string are connected by spaces. The identifier \textit{"<EOS>"} indicates the end of the URL and the start of the request body. The purpose of using this word segmentation method is to preserve words with semantic information as much as possible (like \textit{"SELECT"}). Another benefit is that it makes the result of pattern mining more interpretable. In this case, the pattern we want to obtain is \textit{\{"SELECT", "FROM", "WHERE"\}}. However, if we use n-grams(n=3), the result may be denoted as \textit{\{"SEL", "ELE", ..., "ERE"\}}, which is not intuitive.

\subsubsection{Structure Extraction}
\begin{table}
  \caption{We replace substring with RS(Replacement String) according to its ET(Encoding Type). For strings not belonging to any specially encoded, we substitute each character with RC(Replacement Character) according to its CT(Character Type).}
  \label{tab:freq}
  \begin{tabular}{llll}
    \toprule
    ET & RS & CT & RC \\
    \midrule
    MD5 hash & MD5\_HASH & Arabic Numerals & D\\
    SHA hash & SHA\_HASH & English Alphabet & W \\
    Base64 & BASE64\_ENCODE & Garbage Character & G \\
    Hexadecimal & HEXADECIMAL & Chinese Character& C \\
    Binary & BINARY & Invisible Character& I \\
  \bottomrule
  \end{tabular}
\end{table}
Research observed that in many cases there are "stable" path components that are specific to a particular malicious family or operation \cite{182949}. In other words, there is similarity in the content or structure of the same type of traffic. Inspired by this, the paper extracts structure information based on a series of string substitution rules. In order to cover all the string types that appear in the request data, we have performed frequency statistics. The following Table 1 gives a summary of pivotal rules.

Unlike general natural language, there are plentiful special characters in HTTP request. These special characters usually reflect the traffic characteristics and have high practical value. Therefore, for a substring with length 1, we will not replace it if it is a special character. We traverse other substrings in the segmented string to determine their encoding types (such as md5 or hexadecimal) and convert it to the replacement string according to rules shown in Table 1. If it does not belong to any special encoding form, we traverse each character in the substring, judge what type it is and replace it.
\begin{figure}[htbp]
\includegraphics{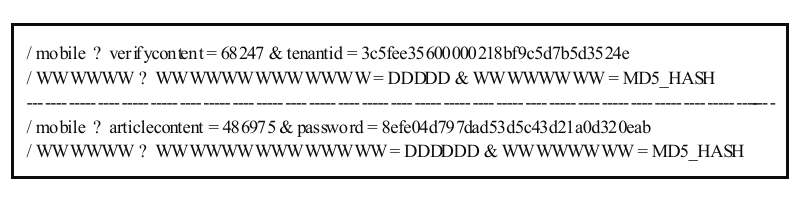}
\caption{An example of structure extraction.}
\end{figure}

An example of structure extraction is shown in Figure 2. Since the encoding type of the substring \textit{"3c5fee35600000218bf9c5d7b5d3524e"} is MD5, we replace it with \textit{"MD5\_HASH"}. For the substring \textit{"68247"}, we replace each number in it with \textit{"D"}, then we can obtain its structure feature, which is denoted as \textit{"DDDDD"}. Obviously, by extracting structural features, we can easily find requests with different content but almost the same structure. Combined content and structural features are beneficial to improve the generalization ability of the model. It is crucial for identifying malicious traffic and discovering malicious patterns.

\section{Our Approach}
\begin{figure*}
\includegraphics[width=14cm,height=8cm]{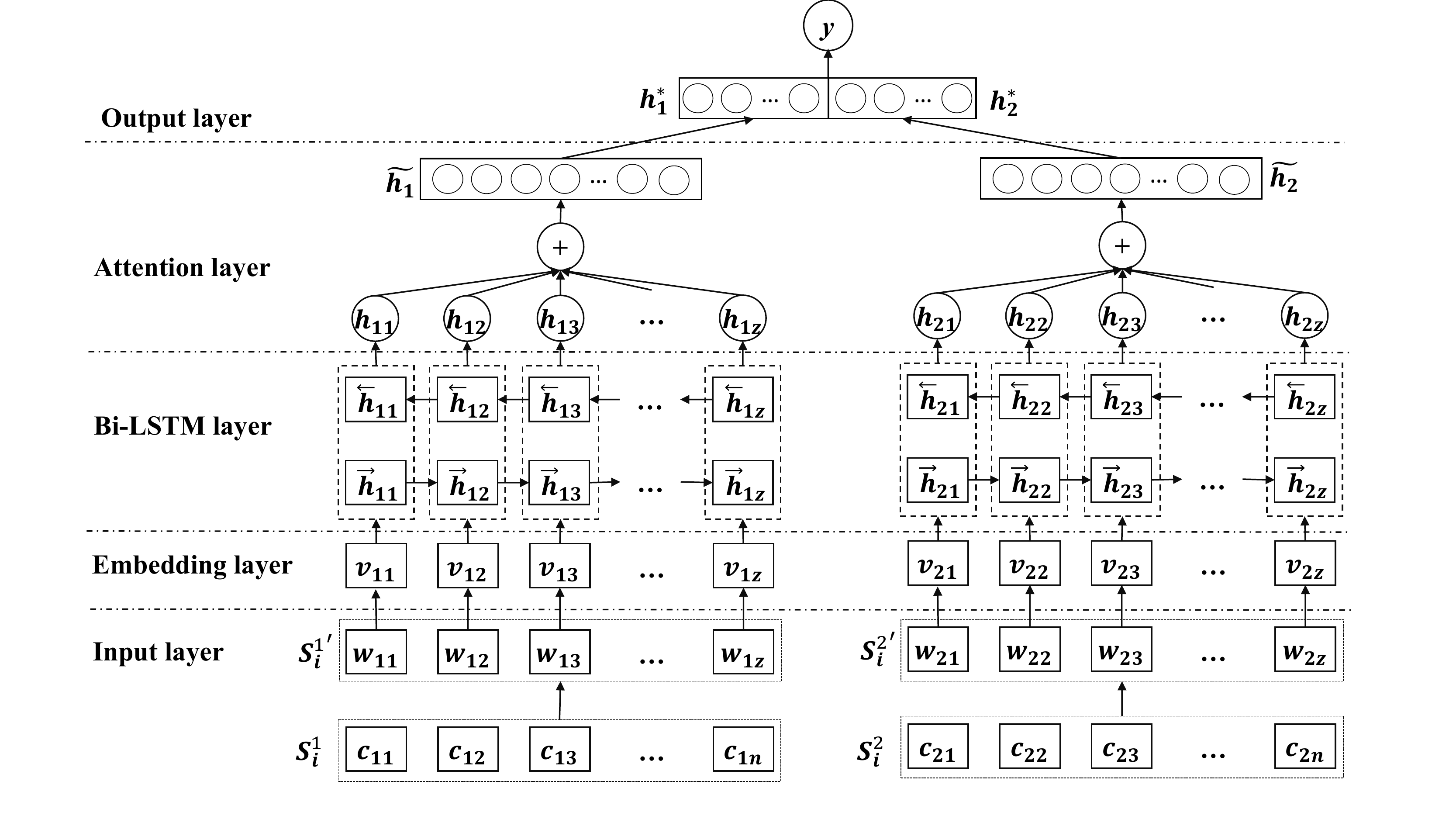}
\caption{Model Architecture.}
\end{figure*}
\subsection{Anomaly HTTP Traffic Detection}
The goal of the proposed algorithm is to identify anomaly HTTP traffic based on semantics and structure of traffic entries. Figure 3 shows the high-level overview of the proposed model. The model based on Bi-LSTM and attention mechanism contains five components: input layer, word embedding layer, Bi-LSTM layer, attention layer and output layer. Before output layer, we train content sequence and structure sequence respectively. Then, we adopt drop-out strategy to avoid overfitting and simply fuse the feature vector of content and structure. In the output layer, we perform classification using a softmax function.

\subsubsection{Problem definition}
Let \begin{math}R=\{R_1,R_2,\cdots,R_N\} \end{math} be the set of HTTP traffic entries after data processing. For each traffic entry \begin{math}R_i(i=1,2,\cdots,N)\end{math}, there are two sequences \begin{math} S_i^1 = \{c_{11}, c_{12}, \cdots, c_{1n}\} \end{math} and \begin{math} S_i^2 = \{c_{21}, c_{22}, \cdots, c_{2n}\}\end{math}, which respectively represent content sequence and structure sequence. Because structure sequence is derived from content sequence, the length of both sequence is equal to n. As a result, we obtained two feature sets to characterize the semantics and structure of traffic.

\subsubsection{Input Layer}
One-hot representation is a widely used and relatively simple word vector generation method in the field of natural language processing. However, the length of the word vector generated by this text preprocessing method is up to vocabulary size. Usually, vectors are usually quite sparse.

In this paper, we use the content and structure sequence after word segmentation as a corpus, and select words that are common in the corpus to build a vocabulary according to term frequency-inverse document frequency (TFIDF). Then, the unique index is generated for each word in the vocabulary. The final input vector with fixed length is composed of indexes. The length of input vector is denoted as \textit{z}, which is a hyper-parameter(the fixed length in this paper is set to 300 because the proportion of sequence length within 300 is 0.8484). The excess part of input sequence is truncated, and the insufficient part is filled with zero. Formally, the sequence of content can be converted to \begin{math} S_i^{\acute{1}}=\{w_{11},w_{12},\cdots,w_{1z}\} \end{math} and the sequence of structure can be expressed as  \begin{math}S_i^{\acute{2}} =\{w_{21},w_{22},\cdots,w_{2z}\}\end{math}.We use a simple example to illustrate the process. Given a sequence of content: \textit{\{'/', 'admin', '/', 'caches', '/', 'error\_ches', '.', 'php' \}}. The input vector with fix length can be denoted as \begin{math} [23,3,23,56,23,66,0,0,\cdots,0] \end{math}, the index of \textit{'admin'} in vocabulary is 3. Since the length of this sequence is less than fixed length, the rest of the vector is filled with zeros.

\subsubsection{Embedding Layer}
Take a content sequence of i-th traffic entry as an example. Given \begin{math} S_i^{\acute{1}}=\{w_{11},w_{12},\cdots,w_{1z}\} \end{math}, we can obtain vector representation \begin{math} v_{1k}\in R^m \end{math} of each word \begin{math} w_{1k}\in R^1(k=1,2,\cdots,z) \end{math}as follows:
\begin{equation}
  v_{1k}=ReLU(W_ew_{1k}+b_e)
\end{equation}
where \begin{math} m \end{math} is the size of embedding dimension, \begin{math} W_e\in R^{m\times1} \end{math}is the weight matrix, and \begin{math} b_e\in R^m \end{math}is the bias vector. ReLU is the rectified linear unit defined as \begin{math} ReLU(v) =max(v, 0) \end{math}, where \begin{math}max()\end{math} applies element-wise to vector.

\subsubsection{Bidirectional Long Short-Term Memory}
We employ Bidirectional Long Short-Term Memory (Bi-LSTM), which can exploit information both from the past and the future to improve the prediction performance and learn the complex patterns in HTTP requests better. A Bi-LSTM consists of a forward and backward LSTM. Given embedding vector \begin{math} \{v_{11},v_{12},\cdots,v_{1z}\} \end{math} of content sequence of i-th traffic entry \begin{math}R_i\end{math}, the forward LSTM \begin{math}\overset{\rightarrow}{f}\end{math}reads the input sequence from \begin{math}v_{11}\end{math} to \begin{math}v_{1z}\end{math}, and calculates a sequence of forward hidden states \begin{math}(\overset{\rightarrow}{h_{11}}, \overset{\rightarrow}{h_{12}},\cdots,\overset{\rightarrow}{h_{1z}})\end{math}(\begin{math} \overset{\rightarrow}{h_{1i}}\in R^p\end{math}and \begin{math}p\end{math} is the dimensionality of hidden states). The backward LSTM \begin{math}\overset{\leftarrow}{f} \end{math}reads the input sequence in the reverse order and product a sequence of backward hidden states \begin{math}(\overset{\leftarrow}{h_{11}},\overset{\leftarrow}{h_{12}},\cdots,\overset{\leftarrow}{h_{1z}})\end{math}(\begin{math} \overset{\leftarrow}{h_{1i}}\in R^p\end{math}). The final latent vector representation \begin{math}h_{1i}=[\overset{\rightarrow}{h_{1i}}; \overset{\leftarrow}{h_{1i}}]^T (h_{1i}\in R^{2p})\end{math} can be obtained by concatenating the forward hidden state \begin{math} \overset{ \rightarrow}{h_{1i}}\end{math} and the backward one \begin{math} \overset{ \leftarrow} {h_{1i}} \end{math}. We deal with the embedding vector of structure sequence in the same way.

\subsubsection{Attention Layer}
In this layer, we apply attention mechanism to capture significant information, which is critical for prediction. The function we use to capture the relationship between \begin{math}h_t\end{math} and \begin{math}h_i(1\leq i<t) \end{math} is called general attention:
\begin{equation}
  \alpha_{ti}=h_t^T W_\alpha h_i
\end{equation}
\begin{equation}
  \alpha_t=softmax([\alpha_{t1},\alpha_{t2},\cdots,\alpha_{t(t-1)}])
\end{equation}
where \begin{math}W_\alpha\in R^{2p\times 2p}\end{math} is the matrix learned by model, \begin{math}\alpha_t\end{math} is the attention weight vector  calculated by softmax function. Then, the context vector \begin{math}c_t\in R^{2p}\end{math} can be calculated based on the weights obtained from Eq.(3) and the hidden states from \begin{math}h_1\end{math} to \begin{math}h_{t-1}\end{math} as follows:
\begin{equation}
  c_t = \sum_{i}^{t-1} \alpha_{ti} h_i
\end{equation}
We combine current hidden state \begin{math}h_t\end{math} and context vector \begin{math}c_t\end{math} to generate the attentional hidden state as follows:
\begin{equation}
  \tilde{h_t}= \tanh(W_c[c_t;h_t])
\end{equation}
where \begin{math}W_c\in R^{r\times 4p}\end{math} is the weight matrix in attention layer, and \begin{math}r\end{math}is the dimensionality of attention state. \begin{math}\tilde{h_1}\end{math} and \begin{math}\tilde{h_2}\end{math}can be obtained using Eq.(2) to Eq.(5), which denote the attention vector of content and structure sequence learned by the model.
\subsubsection{Output Layer}
Before feeding the attention vector into softmax function, the paper apply dropout regularization [16] randomly disables some portion of attention state. It is worth noting that we concatenate vector of content and structure to generate output vector. Vectors incorporating semantic and structural information are eventually used for prediction. The classification probability is calculated as follows:
\begin{equation}
  p= softmax(w_s [h_1^*;h_2^* ]+b_s )
\end{equation}
where \begin{math}h_1^*\end{math}is the output of \begin{math}\tilde{h_1}\end{math}after dropout strategy, \begin{math}h_2^*\end{math}is the output of \begin{math}\tilde{h_2}\end{math}. \begin{math}w_s\in R^{q\times r}\end{math} and \begin{math}b_s\in R^q\end{math}are the parameters to be learned.
\begin{equation}
  \hat{y}=argmax(p)
\end{equation}
where \begin{math}\hat{y}\end{math} is the label predicted by the attention model.
\subsubsection{Objective Function}
The paper calculate the loss for all HTTP traffic entries using the cross-entropy between the ground truth \begin{math}y_i\in(0,1)\end{math} and the predicted \begin{math}p_i(i=1,2,\cdots,N)\end{math}:
\begin{equation}
  L=-\frac{1}{N}\sum_{i=1}^{N} y_i \log(p_{i1})+(1-y_i)\log(1-p_{i1})
\end{equation}
where \begin{math}N\end{math} is the number of traffic entries, \begin{math}p_{i1} \end{math} denotes the probability that the i-th sample is predicted to be malicious.

\begin{figure*}
\includegraphics[width=14cm,height=5cm]{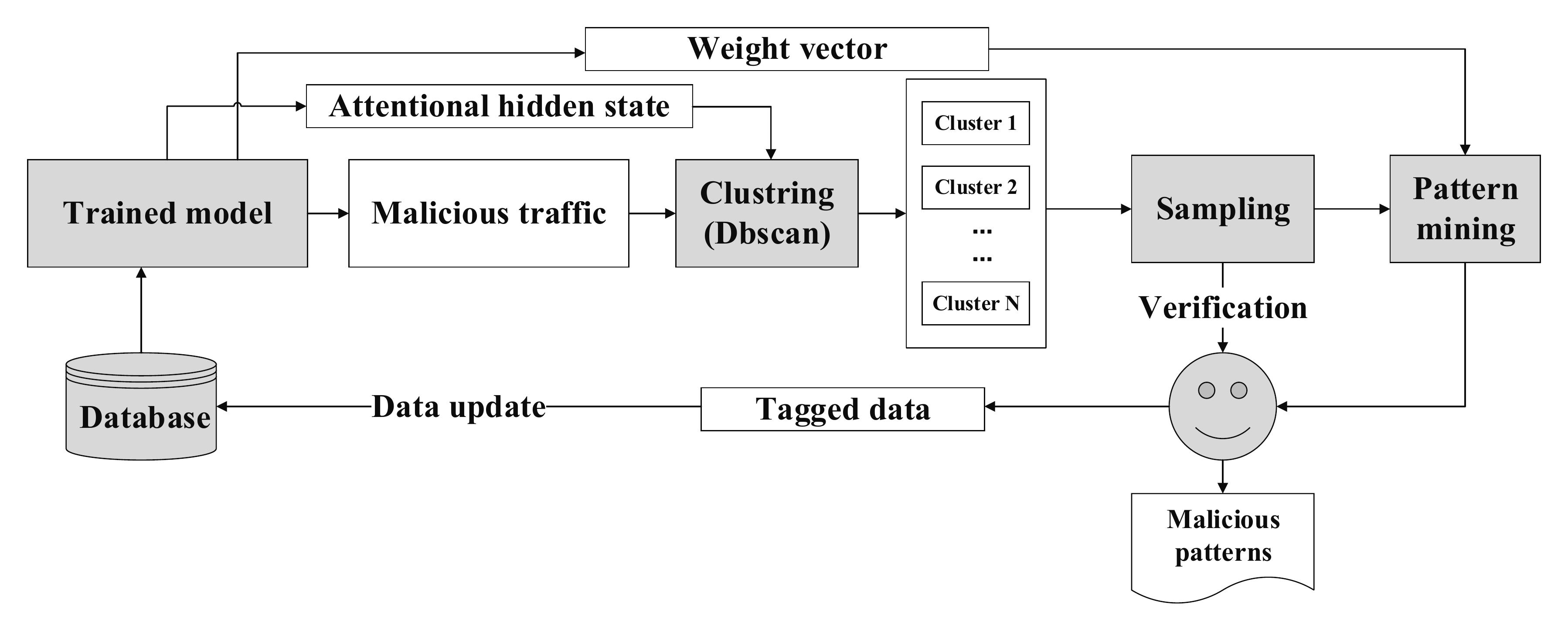}
\caption{The architecture of mining stage.}
\end{figure*}

\subsection{Malicious Pattern Mining}
This section provides in-depth analysis of model results, including malicious pattern mining and verification of model results. Figure 4 shows the architecture of mining stage. The process is explained as follows.

\subsubsection{Clustering}
We cluster traffic entries labeled as malicious by model, which is the basis for validation and pattern mining in our work. Specifically, we use the attentional hidden state extracted from the model as input vector for clustering. The clustering method we apply is DBSCAN \cite{Ester:1996:DAD:3001460.3001507}, a density-based clustering algorithm, which does not require prior declaring the number of clusters. After clustering, we obtain several clusters. Traffic entries in each cluster are similar in content or structure.

\subsubsection{Tag Verification}
In practical applications, massive HTTP traffic requests are generated every day. There is no doubt that manual verification requires a lot of time and efforts. In this paper, we combine clustering and sampling to reduce the workload. After clustering, we extract some samples from each cluster for verification. If the predicted labels of these samples are consistent with the ground truth, then all the prediction results in this cluster are considered correct.

\subsubsection{Malicious Patterns Discovery}
Pattern mining of malicious traffic can help discovering commonalities and characteristics of malicious traffic and generating corresponding rules. Especially for unknown web attacks, analyzing their attack patterns is particularly significant.

As mentioned in section 3.1, the attention weight vector obtained in attention layer can reflect the crucial parts where the model is concerned. Therefore, for each malicious traffic entry, we dig out the key parts according to the corresponding attention weight vector. The greater the weight is, the more important the word is. Then, given a series of HTTP requests in the same cluster, malicious pattern can be summed up by comprehensively analyzing the key parts. The specific steps are as follows:

\textbf{Get Top\_N words as candidate set.} Given a cluster with N traffic requests \begin{math} T=\{t_1,t_2,\cdots,t_N\}\end{math}, we first calculate TFIDF for each word in segmented sequence set to obtain top n words in the entire cluster. The candidate set consisting of these words will be expressed as \begin{math}C=\{c_1,c_2,\cdots,c_n\}\end{math}.

\textbf{Get Top\_m words of each traffic.} Then, we select top m key words \begin{math} K_i=\{k_1,k_2,\cdots,k_m\}\end{math} for each traffic entry \begin{math} t_i(i=1, 2, \cdots, N) \end{math} according to the corresponding weight. As a result, we can obtain a set of key words identified by the model, which can be denoted as \begin{math} K=\{K_1,K_2,\cdots,K_N\}\end{math}.

\textbf{Calculate the co-occurrence matrix.} To discover pattern existed in the cluster, we calculate the co-occurrence matrix of words in candidate set C. The goal is that we want to 	
unearth words that not only frequently occur in this cluster but also recognized by the model as key parts. If we can discovery several words with high attention weight that appear together constantly, then the combination of these words can represent the pattern of such traffic.
\begin{table}
  \caption{Distribution of malicious traffic entries.}
  \label{tab:freq}
  \begin{tabular}{ll}
    \toprule
    Data Type & Number \\
    \midrule
    Deserialization & 6014\\
    CMS & 5836 \\
    File Inclusion & 46438 \\
    SQL Injection & 463776 \\
    Webshell & 288050 \\
    XSS  & 127750 \\
    Sensitive Data Exposure  & 16656 \\
    Middleware Vulnerability  & 47614 \\
    Struts2 Vulnerability  & 42477 \\
    Botnet & 19901 \\
    Total & 1064512 \\
  \bottomrule
  \end{tabular}
\end{table}
\section{Evaluation}
\subsection{Data Set}
Due to the insufficient amount of data in the standard data set, we use real traffic data accumulated over time to validate our approach. Monitoring software is used to capture HTTP traffic packets via the gateway of an international university. The collected data is highly sensitive because it contains most of the network activities during work hours of teachers and students. For these data, we perform manual verification and tagging. In addition, other malicious traffic is collected by vulnerability scans under experimental environment. The total number of labeled data is 2095222, half of them are malicious traffic entries. The types and quantities of malicious samples are shown in the Table 2. Moreover, there is five million unmarked HTTP traffic prepared for model testing.

\subsection{Model Comparison}
\subsubsection{Detection in the Labeled Dataset}
Three baseline methods are used for comparison experiments. Convolutional neural networks (CNNs)\cite{554195} is a class of deep feedforward artificial neural networks, most commonly applied to analyzing visual imagery. So far, it is also widely used in video recognition, recommender systems \cite{NIPS2013_5004} and natural language processing \cite{Collobert:2008:UAN:1390156.1390177}. Recurrent neural networks (RNNs), the collective name for a series of sequence models, which are usually applied to solve sequence problems, such as time series prediction\cite{728168} and speech recognition \cite{DBLP:journals/corr/LiW14a}. Long short-term memory (LSTM), variants of recurrent neural network (RNN), which can avoid the vanishing gradient problem. The proposed model is based on Bi-LSTM, which combines LSTM and bidirectional strategy. It can predict or label each element of the sequence based on the element's past and future contexts.
\begin{table}[htbp]
  \caption{ Model results in the labeled dataset.}
  \label{tab:freq}
  \begin{tabular}{cp{1cm}cp{1cm}cccc}
    \toprule
    Model & Sample Ratio & Precision & Recall & F1-score & AUC \\
    \midrule
    CNN & \multirow{4}{1cm}{1:100} & 0.9637&0.3556&0.5196&0.6778\\
    LSTM & & 0.9408 & 0.6778 & 0.7879 & 0.8387 \\
    Bi-LSTM & & 0.9175 & 0.7448 & 0.8222 & 0.8721 \\
    Our Model& & 0.9561 & 0.9609 & $\textbf{0.9585}$ & $\textbf{0.9795}$ \\
    \midrule
    CNN & \multirow{4}{1cm}{1:10} & 0.8772&0.8239&0.8496&0.9067\\
    LSTM & & 0.9249 & 0.9759 & 0.9497 & 0.9844 \\
    Bi-LSTM & & 0.9866 & 0.9586 & 0.9724 & 0.9787 \\
    Our Model& & 0.9905 & 0.9747 & $\textbf{0.9825}$ & $\textbf{0.9869}$ \\
    \midrule
    CNN & \multirow{4}{1cm}{1:1} & 0.9452&0.9867&0.9656&0.9679\\
    LSTM & & 0.9954 & 0.9947 & 0.9951 & 0.9953 \\
    Bi-LSTM & & 0.9961 & 0.9948 & 0.9955 & 0.9957 \\
    Our Model& & 0.9979 & 0.9963 & $\textbf{0.9971}$ & $\textbf{0.9973}$ \\
  \bottomrule
  \end{tabular}
\end{table}
To illustrate the superiority of the proposed model, we extract positive and negative samples from the data set according to different proportions. The evaluation metrics consist of precision, recall, F1-score and AUC (the area under the receiver operating characteristic curve). Table 3 shows the experimental results of mentioned models. Overall, sequence model is more suitable for traffic detection than convolutional neural network model, especially Bi-LSTM. It is worth mentioning that our model is superior to all baseline models in almost all metrics. In unbalanced data sets, the superiority of the proposed model is even more pronounced.
\subsubsection{Detection in the Unlabeled Dataset}
We conduct comparative experiments in five million unlabeled traffic entries. Traditional filtering rules are utilized to assist model validation. The explanation of the assessment indicators we constructed is as follows:

\textbf{MT\_MODEL.} The number of traffic labeled as malicious by the model.

\textbf{MT\_RULE.} The number of traffic that is labeled as malicious by the model and matches the rules.

\textbf{MT\_NEW.} The number of traffic that is labeled as malicious by the model but does not match the rules.

\textbf{FP.} The number of false positive. (\textit{FP= MT\_MODEL-MT\_RULE-MT\_NEW })

\begin{table}
  \caption{ Model results in the unlabeled dataset.}
  \label{tab:freq}
  \begin{tabular}{ccccc}
    \toprule
    Model & MT\_MODEL & MT\_RULE & MT\_NEW & FP  \\
    \midrule
    CNN & $\textbf{795692}$ & \textbf{208917} & 60973 & 525802 \\
    LSTM & 238689 & 139302 & 14527 & 84860  \\
    Bi-LSTM & 295621 & 191267 & 28753 & 75601 \\
    Our Model& 428270 & $\textbf{216809}$ & $\textbf{150974}$ & $\textbf{60487}$ \\
  \bottomrule
  \end{tabular}
\end{table}
The result of model evaluation in real traffic dataset without label is shown in Table 4. According to MT\_RULE, it can be seen that almost all malicious traffic discovered by rules can be detected by CNN and our model (the number of malicious traffic matching the filtering rules is 217100). Moreover, these models have the ability to discover new malicious traffic. However, MT\_MODEL of CNN is extremely large, which means there are quite a few false positives. Additionally, compared with LSTM and Bi-LSTM, the proposed model is able to disclose more anomalous traffic entries. The FP of our model is the lowest. The result further proves the superiority of our model.

\subsubsection{Model Performance with Different Features}
\begin{figure}
\begin{subfigure}[h]{0.45\linewidth}
\includegraphics[height=3.5cm,width=4.3cm]{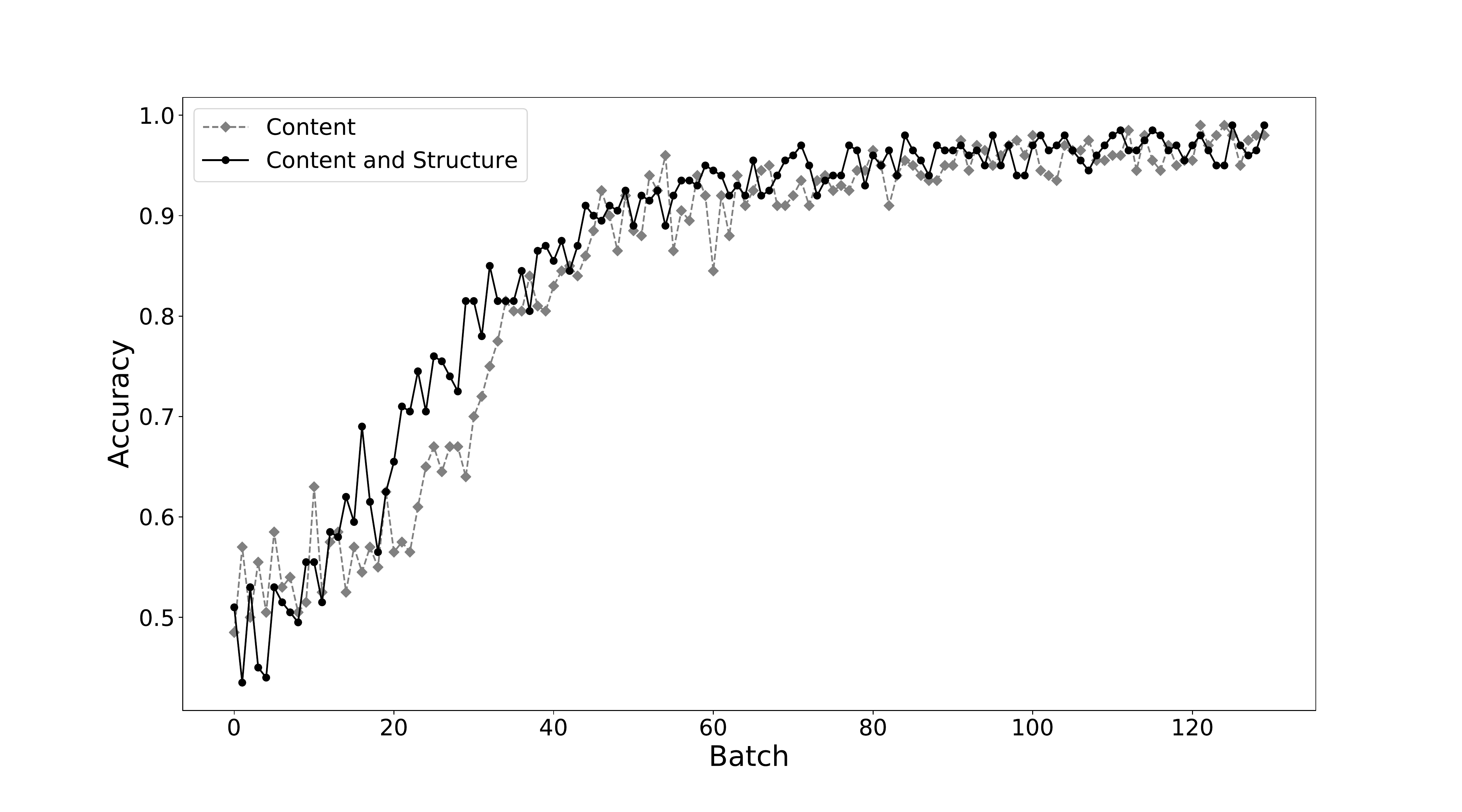}
\caption{Accuracy curve.}
\end{subfigure}
\hfill
\begin{subfigure}[h]{0.45\linewidth}
\includegraphics[height=3.5cm,width=4.3cm]{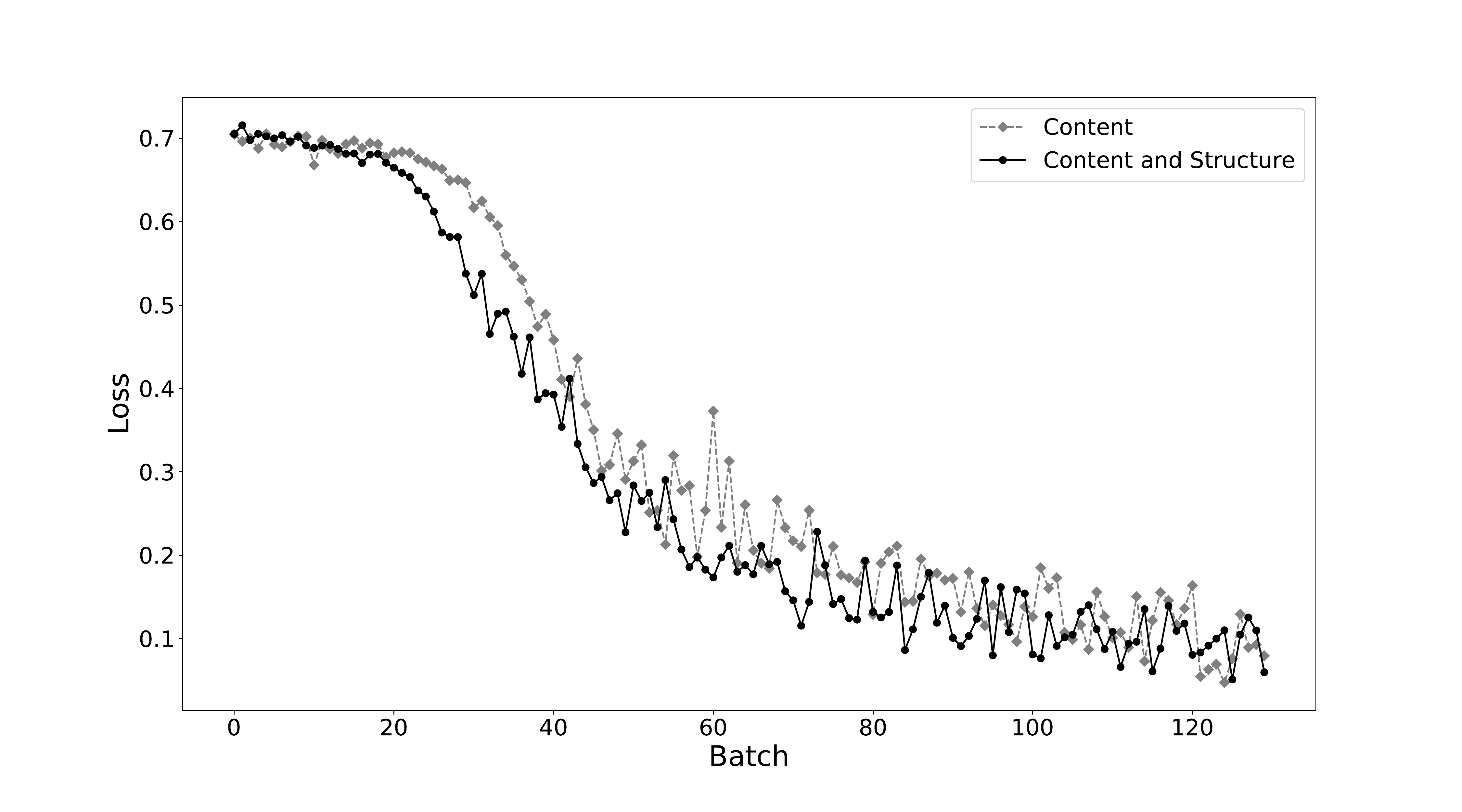}
\caption{Loss curve.}
\end{subfigure}%
\caption{Accuracy curve and loss curve.}
\end{figure}

\begin{figure*}
\begin{subfigure}[h]{0.3\linewidth}
\includegraphics[width=\textwidth]{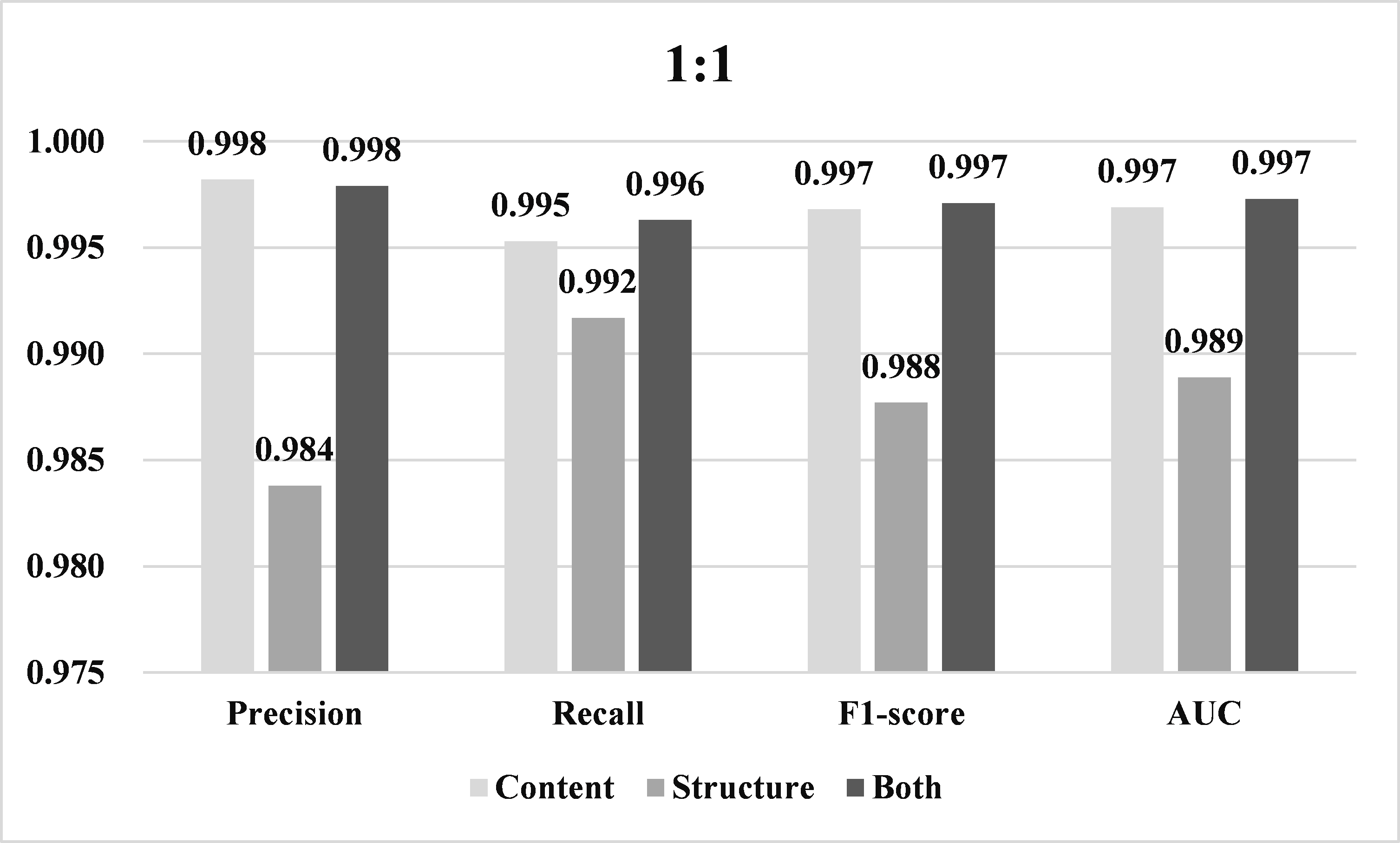}
\caption{The balanced dataset.}
\end{subfigure}
\hfill
\begin{subfigure}[h]{0.3\linewidth}
\includegraphics[width=\textwidth]{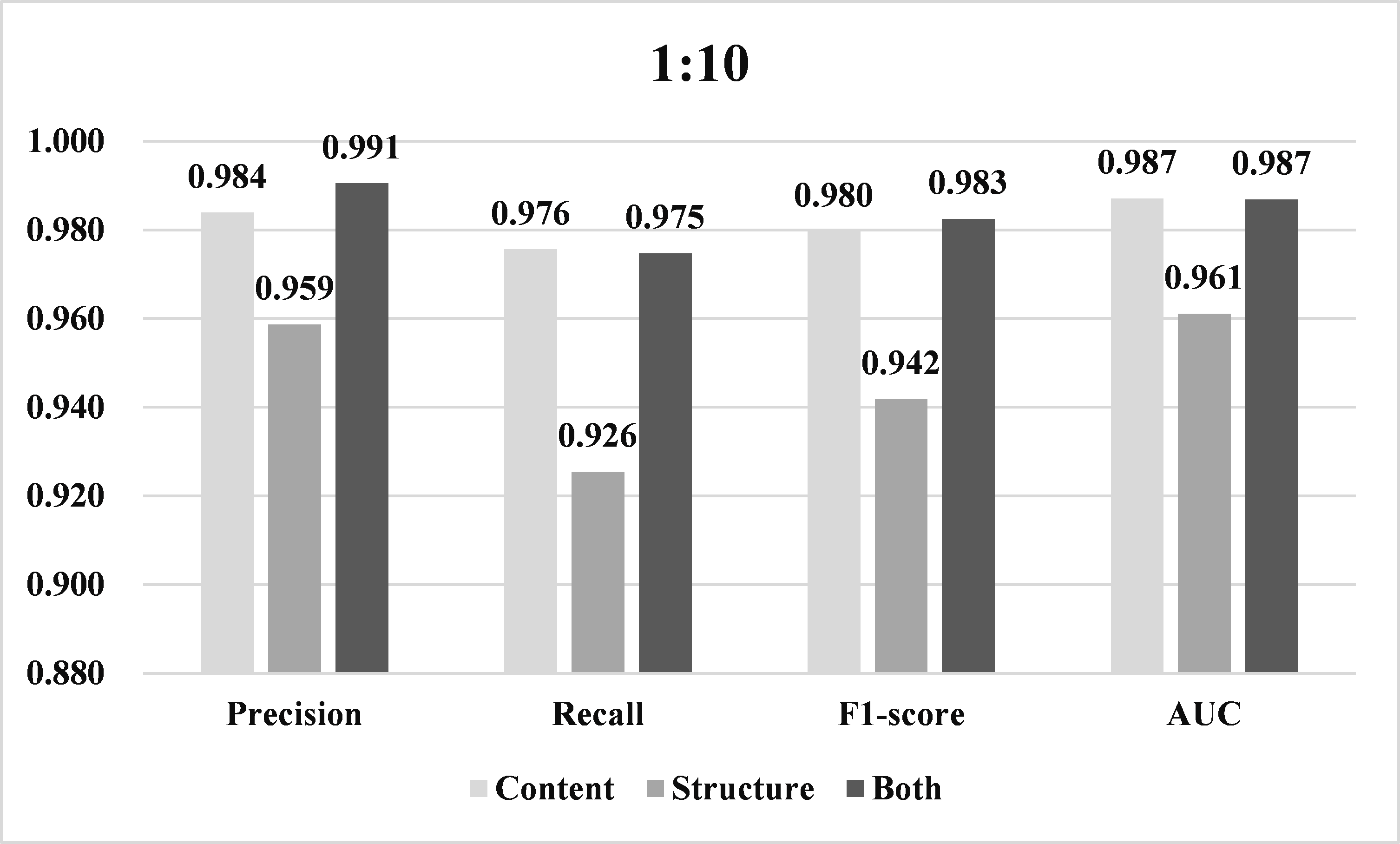}
\caption{The unbalanced dataset(1:10).}
\end{subfigure}
\hfill
\begin{subfigure}[h]{0.3\linewidth}
\includegraphics[width=\textwidth]{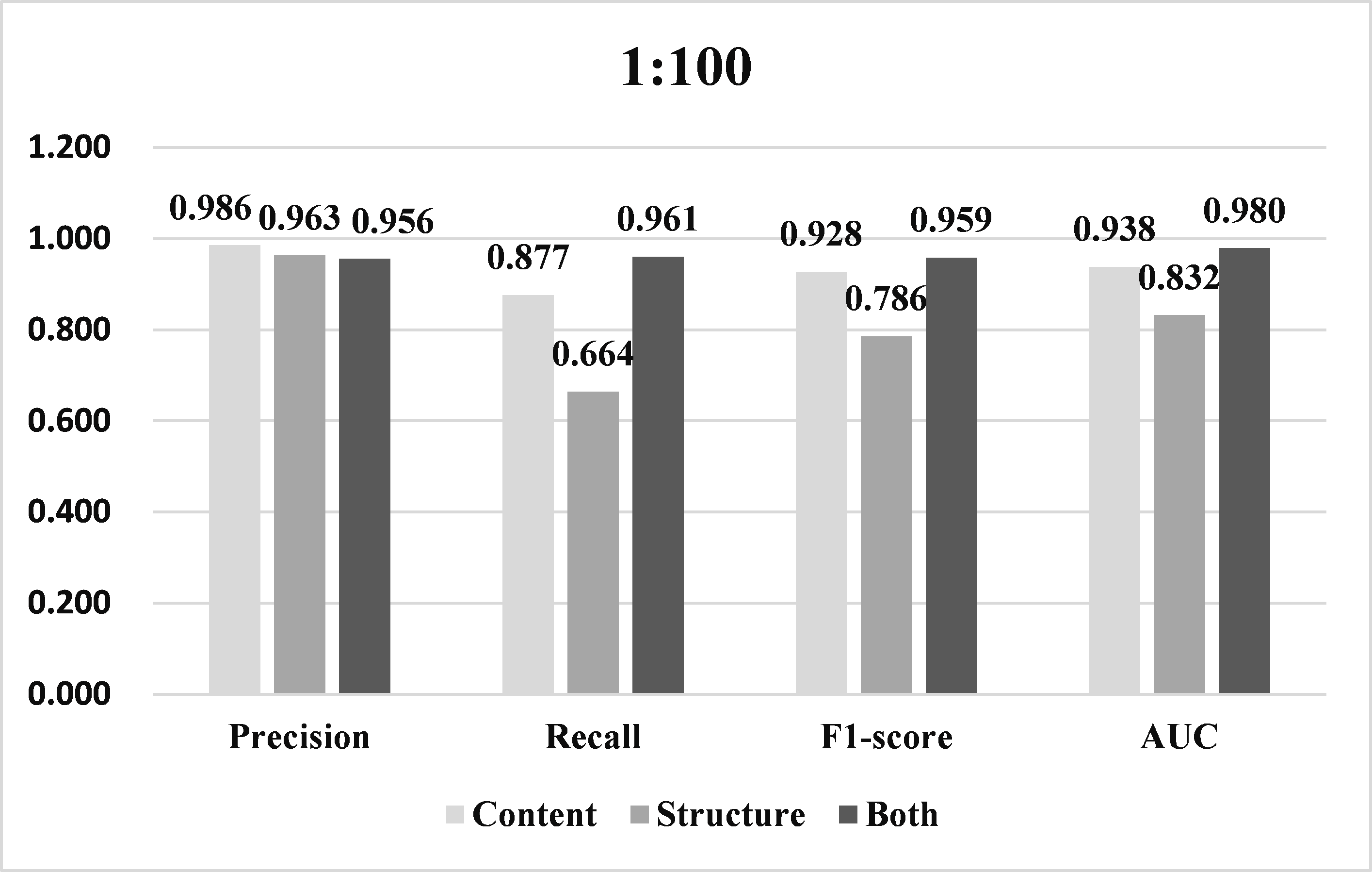}
\caption{The unbalanced dataset(1:100).}
\end{subfigure}%
\caption{Model performance with different features in different datasets.}
\end{figure*}
We record the loss and accuracy of each iteration of the model and draw loss curve and accuracy curve(Figure 5). To balance the memory usage and model training efficiency, the best batch size is set to 200. As we observe from the figure, the model trained based on content and structural features converges faster. In other word, after fusing structural features, the learning rate has been enhanced, and it can reach the convergence state faster.

In order to verify the validity and rationality of the structural features, we first construct three labeled data sets composed of different proportions of positive and negative samples. Then, in these data sets, we use different input features to train the models (including content features, structural features, combined content and structural features). After that, statistical measures of binary classification are used to evaluate the effectiveness of these models. As shown in Figure 6, these models perform better on balanced data sets. The model trained with structure feature is less effective than the other two models in three data sets. After integrating the structural features, the effect of the model has been significantly improved, especially in unbalanced data sets.
\subsubsection{Malicious pattern mining}
In this part, we first cluster anomaly traffic entries to find traffic families. Due to the combination of structural features, the model is able to detect anomalous HTTP requests that are different in content but with similar structure. Some examples of malicious traffic families discovered by the model are shown in Figure 7. It is worth noting that these malicious traffic entries has not appeared in the training set.
\begin{figure}[htbp]
\includegraphics[width=8.5cm, height=6cm]{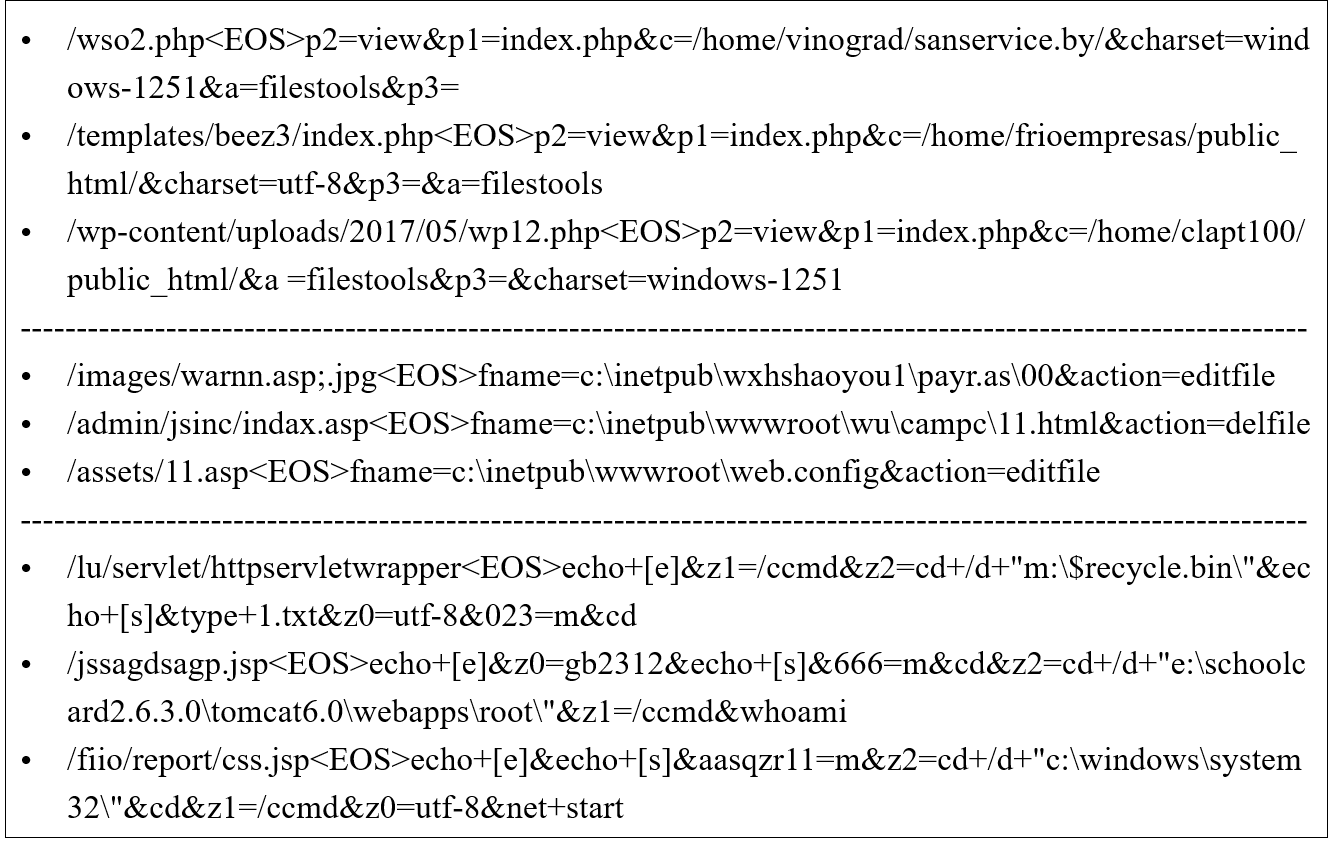}
\caption{Examples of malicious traffic families.}
\end{figure}

As mentioned before, attention weight is pivotal for pattern mining. We can summarize the critical parts of each traffic request according to its weight vector. The visualization of attention vector is shown in Figure 8. The HTTP request is extracted from a cluster after clustering. Color depth corresponds to the weight \begin{math}\alpha_t (Eq.3)\end{math}. Obviously, the key words selected by model are \{\textit{'submit', 'execute', 'wscript', 'action', 'shell'}\}, which are almost in line with the ground truth.
\begin{figure}
\includegraphics[width=8.5cm, height=3cm]{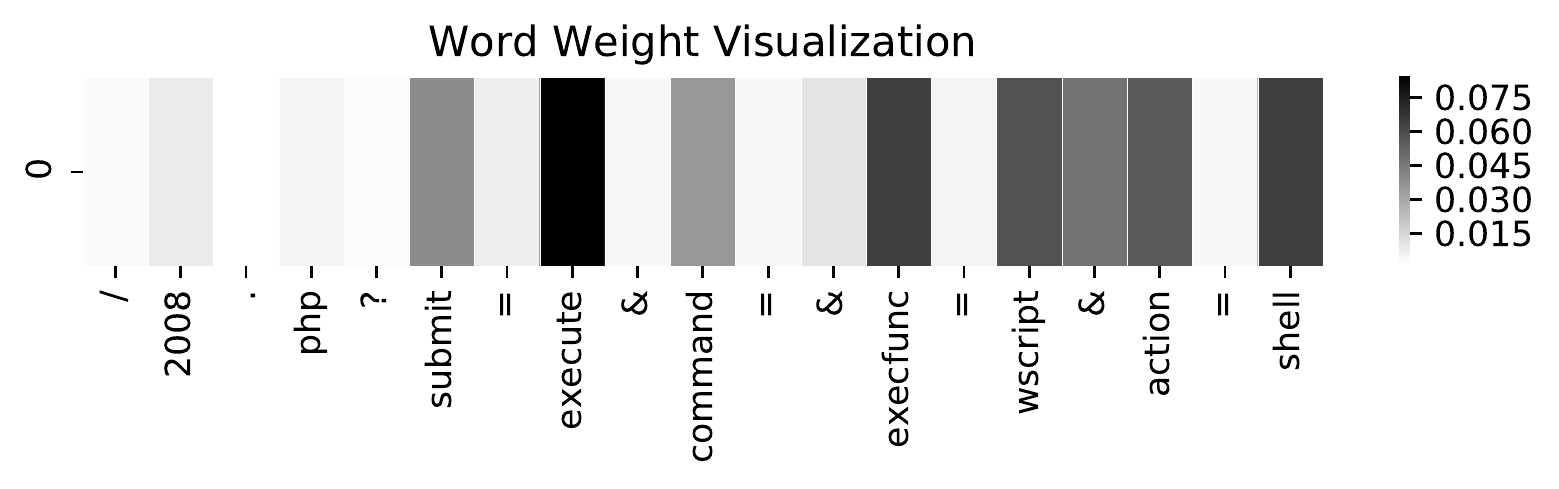}
\caption{Visualization of attention. The depth of the color corresponds to the importance of the word in sentence given by model.}
\end{figure}

To assess the ability of mining patterns, we extract traffic entries by regex matching method to compose the test data set. We use five typical rules and extract one thousand entries for each rule. The specific rules and expressions are shown in Table 5.

\begin{table}
  \caption{ Regex rules.}
  \label{tab:freq}
  \begin{tabular}{ll}
    \toprule
    Regex Rule & Expression \\
    \midrule
    memberaccess.allowstaticmethodaccess & Structs2 \\
    (select.+(from|limit))|(?:(union(.*?)select)) & Sql injection \\
    <(iframe|script|body|img|input) & XSS \\
    \$\_(GET|post|GLOBALS|SERVER)& Webshell \\
    etc.\{0,10\}passwd &File include\\
  \bottomrule
  \end{tabular}
\end{table}

We feed the traffic entries into the proposed model, then the prediction results and attention weights will be received. After that, for each traffic request, we extract top 5 words, top 10 words, and top 20 words ordered by the attention weights and calculate the count of key words appear in top k words (k=5, 10, 20). Just take Structs2 exploitation of vulnerability as an example. According to the first rule, the key words are \textit{"memberaccess"} and \textit{"allowstaticmethodaccess"}. Suppose there are five traffic entries, we collect top 5 words from each traffic according to their weights. If the top 5 words of each traffic contain the key word of malicious traffic (like\textit{"memberaccess"}), the count plus one (we convert the count into a ratio). The experimental results are shown in Table 6. Obviously, for traffic with relatively simple attack patterns, the model can perform well(like \textit{"structs2"} and \textit{"file include"}). Moreover, for some sensitive words (like \textit{"scrip"} and \textit{"limit"}), the model can also well recognize.

\begin{table}
  \caption{Pattern mining evaluation.}
  \label{tab:freq}
  \begin{tabular}{ccccc}
    \toprule
    Rule & Key Words & Top 5 & Top 10 & Top 20\\
    \midrule
    \multirow{2}{1cm}{Rule1} & memberaccess & 0.822 & 0.893 & 0.964 \\
    &allowstaticmethodaccess & 0.697 & 0.850 & 0.923 \\
    \midrule
    \multirow{4}{1cm}{Rule2} & select & 0.744 & 0.832 & 0.871 \\
    &from & 0.502 & 0.650 & 0.792 \\
    &limit & 0.672 & 0.789 & 0.799 \\
    &union & 0.482 & 0.573 & 0.630 \\
    \midrule
    \multirow{5}{1cm}{Rule3} & iframe & 0.721 & 0.817 & 0.821 \\
    &script & 0.789 & 0.820 & 0.854 \\
    &body & 0.340 & 0.554 & 0.590 \\
    &img & 0.470 & 0.334 & 0.652 \\
    &input & 0.456 & 0.562 & 0.652 \\
    \midrule
    \multirow{4}{1cm}{Rule4}&get & 0.523 & 0.585 & 0.789 \\
    &post & 0.434 & 0.687 & 0.697 \\
    &globals & 0.249 & 0.472 & 0.688 \\
    &server & 0.209 & 0.426 & 0.532 \\
    \midrule
    \multirow{2}{1cm}{Rule5}&etc & 0.790 & 0.893 & 0.948 \\
    &passwd & 0.832 & 0.902 & 0.952 \\
  \bottomrule
  \end{tabular}
\end{table}

To further verify the performance of the proposed model on malicious pattern mining, we visualize the results generated by the method mentioned in section 3.2.3 in Figure 9. The darker the color of the square is, the more times the words appear together. The number of co-occurrences of these words can reflect the number of times that they were simultaneously concerned by the model. It is evident that the black area in the upper left corner of the figure is what the model considers important. Hence, the pattern of these traffic can be denoted as \textit{\{"<", "/", "textarea", "script", ">", "alert" \}}.

\begin{figure}[htbp]
\includegraphics[width=8.5cm, height=6cm]{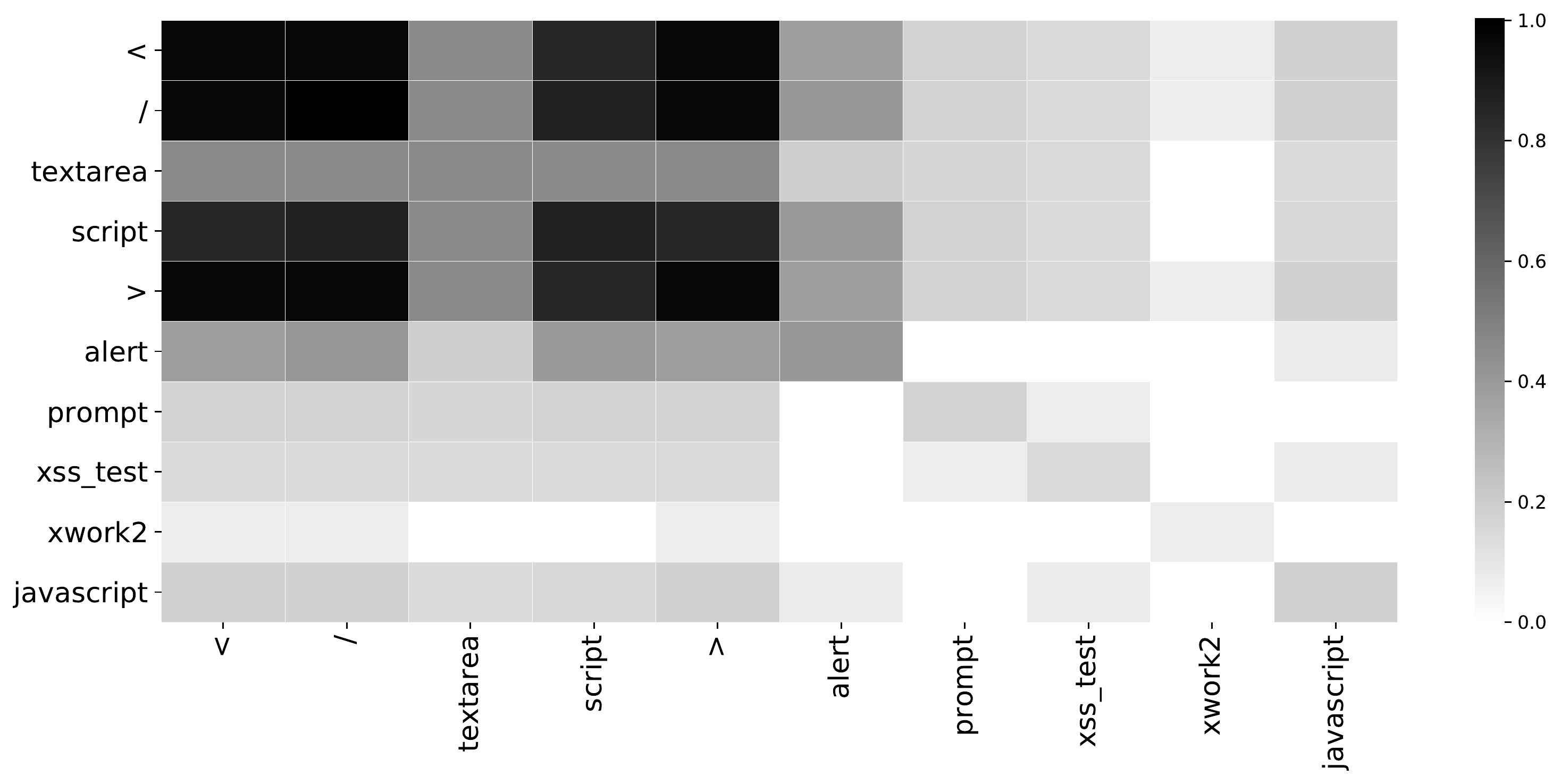}
\caption{visualization of pattern mining.}
\end{figure}

\section{Conclusion}
This paper presents DeepHTTP, a general-purpose framework for HTTP traffic anomaly detection and pattern mining using a deep neural network based approach. We construct a deep neural networks model utilizing Bidirectional Long Short-Term Memory (Bi-LSTM). This enables effective anomaly diagnosis. We design a novel method that can extract the structural characteristics of traffic. DeepHTTP learns content feature and structure feature of traffic automatically, and unearths considerable section of input data. It performs anomaly detection at per traffic entry level, and then mines pattern at cluster level. The intermediate output including attention hidden state and the attentional weight vector can applied to clustering and pattern mining, respectively. By incorporating user feedback, DeepHTTP supports database update and model iteration. Hence it is able to incorporate and adapt to new traffic patterns. Experiments on large traffic datasets have clearly demonstrated the superior effectiveness of DeepHTTP compared with previous methods.

Future works include but are not limited to incorporating other types of deep neural networks into DeepHTTP to test their efficiency. Besides, improving the ability of the model to detect unknown malicious traffic is something we need to further study on in the future.

\section*{Acknowledgement}
This work is partially supported by the National Natural Science Foundation of China (U1736218). 